\documentclass[pra,nobibnotes,showpacs,preprintnumbers,amsmath,amssymb]{revtex4}

\usepackage{graphicx}   
\usepackage{bm}         
\newcommand{\rd}{{\rm d}} 
\newcommand{\re}{{\rm e}} 
\newcommand{\ri}{{\rm i}}

\newcommand{\ak}{\hat{a}^{\phantom\dagger}_{\vec{k}}}
\newcommand{\akd}{\hat{a}^{\dagger}_{\vec{k}}}
\newcommand{\akm}{\hat{a}^{\phantom\dagger}_{-\vec{k}}}
\newcommand{\akmd}{\hat{a}^{\dagger}_{-\vec{k}}}
\newcommand{\az}{\hat{a}^{\phantom\dagger}_{\vec{0}}}
\newcommand{\azd}{\hat{a}^{\dagger}_{\vec{0}}}

\newcommand{\uk}{u_{\vec{k}}}
\newcommand{\kt}{\vec{k'}}
\newcommand{\ukt}{u_{\kt}}
\newcommand{\vkt}{v_{{\kt}}}
\newcommand{\vk}{v_{\vec{k}}}
\newcommand{\Lk}{L_{\vec{k}}}
\newcommand{\bgrund}{|\psi_0\rangle}
\newcommand{\bgrundc}{\langle\psi_0|}
\newcommand{\ord}{{\bf o}}
\newcommand{\ordn}{{\cal O}\left(n\sqrt{n a_0^3}\right)}
\newcommand{\Name}[1]{#1,}
\newcommand{\REVIEW}[4]{{ #1} {\bf #2}, #4 (#3)}
\newcommand{\Book}[1]{{\it #1,}}
\newcommand{\Year}[1]{(#1)}
\newcommand{\Publ}[1]{#1,}
\newcommand{\Page}[1]{#1}

\begin{document}



\title{Ground state energy of a homogeneous
Bose-Einstein condensate beyond Bogoliubov} 
 

\author{Christoph Weiss}
\email{weiss@theorie.physik.uni-oldenburg.de}
\author{
Andr\'e Eckardt}
 
\affiliation{
Institut f\"ur Physik, Carl von Ossietzky Universit\"at, 
                D-26111 Oldenburg, Germany}

\date{\today}

\begin{abstract}
The standard calculations of the ground-state energy of a homogeneous Bose gas rely on
approximations which are physically reasonable but difficult to control.
Lieb and Yngvason [Phys.~Rev.~Lett.~{\bf 80}, 2504 (1998)] have proved rigorously that the commonly
accepted leading order term of the ground state energy is correct in the zero-density-limit. 
Here, strong indications are given that also the next to leading term is correct. 
It is shown that the first terms obtained in   
a perturbative treatment
provide contributions which are lost in the Bogoliubov approach. 
\end{abstract} 

 \maketitle

  
As dilute weakly interacting Bose gases are experimentally realisable since~1995~\cite{exp1,exp2}, there is a renewed interest
in principal results for the ground state energy derived some 50 years ago~\cite{LeeYang57,LeeHuangYang57,BruecknerSawada57,Dyson57,
HugenholtzPines59,Lieb63,Bogoliubov}. In recent years, several subtle issues have been clarified~\cite{BraatenEtAl01,BraatenEtAl99,Leggett03,ChernyShanenko00}.
Consider a weakly interacting translationally invariant Bose gas in the thermodynamic limit
with 
non-negative two-particle interaction potentials excluding bound states of two or more
particles.
In momentum representation, the $N$-particle Hamiltonian in second quantisation 
reads~\cite{Landau00}
\begin{equation}
\label{eq:hsecond}
  \hat{H} = \sum_{\vec{k}}\frac{\hbar^2k^2}{2m}\akd\ak
  +\frac12\sum_{\left\{\vec{k}_i\right\}} 
  \langle\vec{k}_1\vec{k}_2|{U}|\vec{k}_3\vec{k}_4\rangle
  \hat{a}^{\dag}_{\vec{k}_1} \hat{a}^{\dagger}_{\vec{k}_2}
  \hat{a}^{\phantom\dag}_{\vec{k}_3} \hat{a}^{\phantom\dag}_{\vec{k}_4}\;,
\end{equation}
where the matrix elements 
\begin{eqnarray}
\label{eq:matrix}
\langle\vec{k}_1\vec{k}_2|{U}|\vec{k}_3\vec{k}_4\rangle
&=&
\delta_{\vec{k}_1+\vec{k}_2,\vec{k}_3+\vec{k}_4}
\widehat{U}(\vec{k}_2-\vec{k}_4)
\;
\end{eqnarray}
are given by the Fourier transform of the potential
\begin{equation}
  \widehat{U}(\vec{k})\equiv\frac1V\int_{V} U(\vec{r})\re^{\ri\vec{k}\cdot\vec{r}}\,\rd^3r\;.
\label{eq:uhat}
\end{equation}
This approach is valid both for soft potentials (non-pseudo-potentials for which the Fourier
transform exists) and the pseudo-potential usually used to model hard sphere
interaction
\begin{equation}
\label{eq:U0}
U_0(\vec{r})=\frac{4\pi a\hbar^2}m\delta(\vec{r})\frac{\partial}{\partial r}r\;,
\end{equation}
where~$a$ is the hard sphere diameter.

For the pseudo-potential~(\ref{eq:U0}), neglecting terms of the Hamiltonian which are
believed to be small leads to
 the Lee-Yang formula~\cite{LeeYang57,LeeHuangYang57} for
the ground-state energy per particle~$e_0\equiv{E_0}/N$ 
\begin{equation}
  e_0
=
 \frac{2\pi a \hbar^2}{m} n
   \left[ 1 + \frac{128}{15\sqrt{\pi}} \sqrt{n a^3} 
+\ord\left(\sqrt{n a^3}\right)\right] \; ,
\label{eq:e0}
\end{equation}
where~$n\equiv N/V$ is the density. The~$\ord\left(\sqrt{n a^3}\right)$ includes 
a term proportional to~$na^3\ln\left(na^3\right)$ as well as higher order terms (see
ref.~\cite{HugenholtzPines59} and 
references therein) which are very small for realistic experimental
values of~$na^3$~\cite{Leggett01}.
According to the so-called ``Landau postulate'', eq.~(\ref{eq:e0}) is
the ground state energy for any non-negative interaction
potential if the hard sphere diameter~$a$ is replaced by the s-wave
scattering length~\cite{LandauLifshitz02, Leggett01}. This is supported by the fact that
known potential dependent corrections to~$e_0$ are given by~${Cn a^3}{2\pi a \hbar^2} n/m$~\cite{HugenholtzPines59}. The constant~$C$ has only recently
been calculated explicitly by Braaten {\it et al.} in terms of a quantity defined by the $3\to3$ scattering
amplitude~\cite{BraatenEtAl99,BraatenEtAl01}. 

Lieb {\it et al.\/} \cite{Lieb98,Lieb02} have calculated rigorous bounds to prove that in the
limit of vanishing
density~$n$, the leading order term of the ground state of the complete Hamiltonian is indeed
given by the leading term of the Lee-Yang formula:
\begin{equation}
\label{eq:bound_lieb}
 1-C(na^3)^{1/17} \le \frac {e_0}{\frac{2\pi a \hbar^2}{m} n}
 \le
1+11.3(na^3)^{1/3}\;,\hspace{1cm}C>0\;.
\end{equation}
 However, despite
many efforts, the next to leading term could not be proved rigorously;  neither its sign nor
the  exponent of~$na^3$ has been established beyond doubt.

A standard
textbook derivation of equation~(\ref{eq:e0}) uses the Bogoliubov approximation to replace the
quartic Hamiltonian~(\ref{eq:hsecond}) by a quadratic Hamiltonian which can be diagonalised
exactly with the help of the Bogoliubov transformation. Repeating the same analysis for soft
potentials leads to the formula of Brueckner 
and Sawada~\cite{BruecknerSawada57,ChernyShanenko00,Weiss03}
\begin{equation}
\label{eq:e0bs}
  e_0^{(0)} =
  \frac{2\pi \hbar^2 (a_0+a_1)}{m} n
+\frac{2\pi \hbar^2 a_0 }{m} \frac{128}{15\sqrt{\pi}} n\sqrt{n a_0^3}\;,
\end{equation}
where~$a_0$ and~$a_1$ are the first and second Born approximation of the scattering length.
In general, the s-wave scattering length~$a$ is not identical with the sum of the first two
Born approximations~$a_0+a_1$.  Thus, the
terms of the Hamiltonian neglected in the Bogoliubov approximation have to contribute in leading
order in the density to the ground state energy per particle.

In this letter we use quantum mechanical perturbation theory to show how the missing terms of
the Born series can be recovered. Furthermore, for very soft potentials, already
first order perturbation theory indicates that the 
next to leading term of the ground state energy of the full Hamiltonian 
is also given by the term of the Lee-Yang formula.
If one accepts the Landau postulate, any experimentally realistic interaction potential can
be replaced by a soft potential with a rapidly converging Born series. However, our approach
is not restricted to such potentials; an
 extension to non-regular potentials like the pseudo-potential~(\ref{eq:U0}) will be given in a
subsequent paper.


We also improve the upper bound~(\ref{eq:bound_lieb}) for the ground state energy. 
Lieb {\it et al.\/} used a product-ansatz for the
$N$-particle wave function to calculate their upper bound according to the variational
theorem. 
Here, we use the Bogoliubov ground state (which is the exact
quantum-mechanical ground state of an
approximate Hamiltonian) to calculate an 
upper bound on the ground state energy of the complete Hamiltonian.


Although we do not use the Born
approximation for the scattering length at any point in our calculations, several expressions
can be identified with terms of the Born series which are derived in a first step. Before we
can calculate the upper
bound, we describe the properties of the Bogoliubov ground state for soft potentials.

\section{Born series}
Often, the 
scattering length is approximated by the first term of the Born series 
\begin{equation}
  a_0
  =
  \frac m{4\pi\hbar^2}\int U(\vec{r})\, \rd^3 r \;.
\end{equation}
The next two terms of the Born series are given by~\cite{CT}
\begin{eqnarray}
\label{eq:a1}
  a_1
  &=&
-
  \left(\frac m{4\pi\hbar^2}\right)^2
  \int U(\vec{r}) \int\frac{U(\vec{r_1})}{\left|\vec{r}-\vec{r}_1\right|}
   \, \rd^3 r_1 \, \rd^3 r 
\\
  a_2
  &=&
  \left(\frac m{4\pi\hbar^2}\right)^3
  \int U(\vec{r})\left[\int \frac{U(\vec{r}_1)}{\left|\vec{r}-\vec{r}_1\right|}
\left(
\int \frac{U(\vec{r}_2)}
{\left|\vec{r}_1-\vec{r}_2\right|}
\, \rd^3 r_2 \,\right) \rd^3 r_1 \right]\, \rd^3 r \;;
\end{eqnarray}
higher order terms can be calculated analogously.
The identity~\cite{Huang57}
\begin{equation}
 \frac1{\left|\vec{r}-\vec{r}_1\right|}
=
\frac{4\pi}V\sum_{\vec{k}\ne\vec{0}}
\frac{\re^{\ri\vec{k}\cdot(\vec{r}-\vec{r}_1)}}{\vec{k}^{\,2}}+{\cal O}(V^{-1/3})
\end{equation}
is useful to rewrite the terms
of the Born series to 
expressions in momentum space.
Using eq.~(\ref{eq:uhat}), we obtain both the well known formula~(see {\it e.g.\/} ref.~\cite{LandauLifshitz02})
\begin{equation}
a_1 = {-}
  \frac {V}{4\pi}\left(\frac m{\hbar^2}\right)^2
\sum_{\vec{k}\ne\vec{0}}\frac{\left(\widehat{U}(\vec{k})\right)^2}{{\vec{k}}^{\,2}}
\end{equation}
and
\begin{equation}
a_2 =  \frac {V}{4\pi}\left(\frac m{\hbar^2}\right)^3
\sum_{\vec{k}_1,\,\vec{k}_2\ne\vec{0}}
\frac{\widehat{U}(\vec{k}_1)\widehat{U}(\vec{k}_2)
\widehat{U}({\vec{k}_1\!-\!\vec{k}_2)}}
{\vec{k}_1^{\,2}\vec{k}_2^{\,2}}\,.
\end{equation}


\section{ Bogoliubov ground state}
For any (normalised) $N$-particle wave-function~$\bgrund$, the variational theorem yields an
upper bound on the ground state energy per particle~$e_0$ of the complete Hamiltonian~(\ref{eq:hsecond})
\begin{equation}
  e_0\le\frac 1N\bgrundc \hat{H}\bgrund\;.
\end{equation}
Here, we use the Bogoliubov ground state as our~$\bgrund$. The idea of the
Bogoliubov transformion is to express the operators $\hat{a}^{\dag}_{\vec{k}}$ and
$\hat{a}^{\phantom\dag}_{\vec{k}}$ which create or annihilate particles with
momentum~$\vec{k}$ by quasi-particle operators $\hat{b}^{\dag}_{\vec{k}}$ and
$\hat{b}^{\phantom\dag}_{\vec{k}}$:
\begin{eqnarray}
   \hat{a}_{\vec{k}} & = & \uk \hat{b}_{\vec{k}}^{\phantom\dag} - \vk
\hat{b}_{-\vec{k}}^\dagger\;, \quad\vec{k}\ne\vec{0}\;,
 \\
\hat{a}_{\vec{0}}&=&\hat{b}_{\vec{0}}
\label{BBDG}
\end{eqnarray}
with
\begin{equation}
   [ \hat{b}^{\phantom\dag}_{\vec{k}_1} , \hat{b}_{\vec{k}_2}^\dagger ] =
\delta_{\vec{k}_1,\vec{k}_2}\;,\quad
[ \hat{b}^{(\dag)}_{\vec{k}_1} , \hat{b}_{\vec{k}_2}^{(\dagger)} ] =0
\;.  
\label{BOCO}
\end{equation}
Without the unnecessary approximation~$\widehat{U}(\vec{k})\approx\widehat{U}(\vec{0})$
(which leads to 
unphysical divergences in both the Born series and in
the calculations for the ground state energy~\cite{LandauLifshitz02,Weiss03}),
the (real) transformation parameters for isotropic interaction potentials (for
which~$\widehat{U}(\vec{k})=\widehat{U}(-\vec{k})$) are defined via: 
\begin{eqnarray}
   \uk &=& \frac{1}{\sqrt{1 - \Lk^2}}\;,\vspace{1cm} 
   \vk 
=
\frac{\Lk}{\sqrt{1 - \Lk^2}} \; ,\\
 \Lk &=& 1 + \frac{\frac{\hbar^2\vec{k}^2}{2m} - \varepsilon(\vec{k})}
                  {N \widehat{U}(\vec{k})} \; ,
\\
\varepsilon(\vec{k}) &=&\left[\left(\frac{\hbar^2\vec{k}^2}{2m}\right)^2
+N\widehat{U}(\vec{k})\frac{\hbar^2\vec{k}^2}{m}
\right]^{1/2}\;.
\label{eq:eps}
\end{eqnarray}

In the Bogoliubov ground state~$\bgrund$ on average $N_0$ particles are in the ground state
characterised by~$\vec{k}=\vec{0}$. 
The macroscopically occupied ground state allows a grand-canonical description
of the excited states.
One has
\begin{eqnarray}
   \bgrundc\hat{b}_{\vec{0}}^\dag
   \hat{b}_{\vec{0}}^{\phantom\dag}\bgrund &=&  {N}_0\;,
\label{eq:b0_1}
\\
   \bgrundc\hat{b}_{\vec{0}}^{\phantom\dag}
   \hat{b}_{\vec{0}}^{\phantom\dag}\bgrund &=&  {N}_0\;,
\label{eq:b0_2}
\\
\label{eq:b0_3}
   \hat{b}_{\vec{k}}^{\phantom\dag}\bgrund &=&0\;\quad\mbox{for}\quad\vec{k}\ne\vec{0}\;,
\end{eqnarray}
where the first equation defines~$ {N}_0$ whereas the second
 is only true in the thermodynamic limit when~$N_0\to\infty$
as~$N\to\infty$  (such that the error in the approximation~$\sqrt{N_0+1}\simeq\sqrt{N_0}$ is
negligible).
It relies on the fact that the number of particles in the ground state fluctuates for a
weakly interacting Bose gas even at zero temperatures. 
The contribution to the total wave
function of wave functions with
exactly~\mbox{$N_0\!+\!\nu$} or exactly \mbox{$N_0\!+\!\nu\!+\!1$} particles are practically the same. However, we would like
to stress at this point that the calculation of an upper bound does not use special
assumptions  about the ground state --- it is of course possible to construct a wave function
which satisfies eqs.~(\ref{eq:b0_1}) to~(\ref{eq:b0_3}). A wave function with a phase factor in eq.~(\ref{eq:b0_2}) or
even with $\bgrundc \hat{b}_{\vec{0}}^{\phantom\dag}\hat{b}_{\vec{0}}^{\phantom\dag}\bgrund=0$
would lead to a higher upper bound and thus can be discarded here.

\section{ Calculating the upper bound} Let us now write the full
Hamiltonian~(\ref{eq:hsecond}) in the form
\begin{equation}
 \hat{H} = \hat{H}_{\rm I}+\hat{H}_{\rm II}+\hat{H}_{\rm III}+\hat{H}_{\rm IV},
\end{equation}
where the first part includes both the kinetic energy and those terms with all
four~$\vec{k}_i=\vec{0}$, the second those contributions with two momenta equal to zero. In the third
part, $\hat{H}_{\rm III}$, only one~$\vec{k}_i$ vanishes whereas in the last all are different from
zero:
\begin{eqnarray}
 \hat{H}_{\rm I} &=& 
\frac12\widehat{U}(\vec{0})\azd\azd\az\az
+\sum_{\vec{k}}\frac{\hbar^2k^2}{2m}\akd\ak
\\
\hat{H}_{\rm II} &=&\frac12\sum_{\vec{k}\ne\vec{0}} \nonumber
\left\{\widehat{U}(\vec{k})\akd \akmd\az\az + \widehat{U}(\vec{k})\azd\azd\ak\akm\right.
+\widehat{U}(\vec{k})\azd\akd\ak\az
\\\nonumber
&&\quad\quad+
\left.\widehat{U}(\vec{k})\akd\azd\az\ak\right.
+
\left.
\widehat{U}(\vec{0})\akd\azd\ak\az+\widehat{U}(\vec{0})\azd\akd\az\ak
\right\}
\\
\hat{H}_{\rm IV} &=&
\frac12\sum_{\left\{\vec{k}_i\ne\vec{0}\right\}} 
  \langle\vec{k}_1\vec{k}_2|{U}|\vec{k}_3\vec{k}_4\rangle
  \hat{a}^{\dag}_{\vec{k}_1} \hat{a}^{\dagger}_{\vec{k}_2}
  \hat{a}^{\phantom\dag}_{\vec{k}_3} \hat{a}^{\phantom\dag}_{\vec{k}_4}\;.
\end{eqnarray}
We have
\begin{equation}
\bgrundc\hat{H}_{\rm III}\bgrund=0
\end{equation}
as the total number of creation and annihilation operators for~$\vec{k}\ne\vec{0}$ is
odd. Thus, the precise equation for~$H_{\rm III}$ is not relevant here.

All sums 
are expanded in powers of~$n$ with
the following procedure: if $\left(\frac{\partial}{
\partial n}\right)^{\nu}\sum\ldots$ is finite and the next derivative with respect to~$n$
diverges, the expression~$\frac{\partial}{\partial\sqrt{n}}\left(\frac{\partial}{\partial
n}\right)^{\nu}\sum\ldots$ is evaluated. In the thermodynamic limit, the resulting integrals
can be evaluated for low densities after the substitution~${\hbar^2k^2}\equiv
2mN\widehat{U}(\vec{0})y^2$~\cite{ChernyShanenko00,Weiss03}. To calculate the depletion~$N-N_0$, we
also apply the fact that 
a complete set of eigenfunctions was used to obtain the second quantised Hamiltonian in
the form~(\ref{eq:hsecond}). Thus, the conservation of the total number of particles~$N$ can be
expressed as
\begin{equation}
  N=\bgrundc \hat{a}^{\dag}_{\vec{0}}\hat{a}^{\phantom\dag}_{\vec{0}}\bgrund
+\sum_{\vec{k}\ne\vec{0}}\bgrundc \hat{a}^{\dag}_{\vec{k}}\hat{a}^{\phantom\dag}_{\vec{k}}\bgrund\;.
\end{equation}
Because of~$\hat{a}_{\vec{0}}=\hat{b}_{\vec{0}}$ 
and~$\bgrundc \hat{b}^{\dag}_{\vec{0}}\hat{b}^{\phantom\dag}_{\vec{0}}\bgrund=N_0\,$, this leads
to
\begin{equation}
\label{eq:depletion}
   N-N_0=\sum_{\vec{k}\ne\vec{0}}
\vk^2=N\frac8{3\sqrt{\pi}}\sqrt{na_0^3}\;.
\end{equation}
The fact that in this formula for the
depletion, $a_0$ rather than $a$ appears, is due to the approximate ground state
wave-function used at this point. 

Even for the anomalous condensate fluctuations of a homogeneous Bose
gas~\cite{Weiss97,Giorgini98,Meier99} we can 
replace~$\bgrundc\azd\azd\az\az \bgrund/N$ by~$\bgrundc\azd\az \bgrund^2/N\,$ for large~$N$. 
 We get including the order~$\ordn$
\begin{eqnarray}
\frac{\bgrundc\hat{H}_{\rm I}\bgrund}N
= 
\frac{2\pi\hbar^2a_0}m  n-
\frac{2\pi\hbar^2 a_1}mn 
-\frac{2\pi\hbar^2 a_0}m n\frac{16}{3\sqrt{\pi}}\sqrt{na_0^3}
-\frac{2\pi\hbar^2 a_0}mn\frac{64}{5\sqrt{\pi}}\sqrt{a_0^3n}
\end{eqnarray}
 and
\begin{eqnarray}
\label{eq:h2}
\frac{\bgrundc\hat{H}_{\rm II}\bgrund}N
= 
2\frac{2\pi\hbar^2a_1}m  n+
\left(\frac{2\pi\hbar^2 a_0}mn\frac{16}{\sqrt{\pi}}
- 
\frac{2\pi\hbar^2 a_1}mn\frac{16}{3\sqrt{\pi}}
+
\frac{2\pi\hbar^2 a_0}mn\frac{32}{3\sqrt{\pi}}\right)\sqrt{a_0^3n} \;.
\end{eqnarray}
The only non-zero contributions to~$\frac1N\bgrundc\hat{H}_{\rm IV}\bgrund$ 
come from expressions
where at least two~$\vec{k}_i$ have the same modulus. We find:
\begin{eqnarray}
\frac{\bgrundc\hat{H}_{\rm IV}\bgrund}N
&=&
\frac1N\sum_{\vec{k},\,\kt\ne\vec{0}}
\frac12\vk\uk\vkt\ukt\widehat{U}(\vec{k}-\kt) 
=
\frac{2\pi\hbar^2}m a_2 n
+ \frac{16}{\sqrt{\pi}}\frac{2\pi\hbar^2}m a_1n\sqrt{na_0^3}
\end{eqnarray}
as the sums 
$
\frac1N\sum_{\vec{k},\,\kt\ne\vec{0}}
\frac12\vk^2\vkt^2\widehat{U} (\vec{0})
$
and 
$ 
\frac1N\sum_{\vec{k},\,\kt\ne\vec{0}}
\frac12\vk^2\vkt^2\widehat{U}(\vec{k}\!-\!\kt) 
$ 
are of order~$n^2a_0^3$.
Thus, including orders of~$\ordn$
 we get the upper bound~${\bgrundc\hat{H}\bgrund}/N$: 
\begin{eqnarray}
\label{eq:newbound}
e_0
\le
\frac{2\pi\hbar^2(a_0+a_1+a_2)}mn
+
\frac{2\pi \hbar^2a_0}{m} n
   \frac{128}{15\sqrt{\pi}} \sqrt{n a_0^3}
+ 
\frac{2\pi \hbar^2a_1}{m} n
   \frac{32}{3\sqrt{\pi}} \sqrt{n a_0^3}\;,
\end{eqnarray}
which lies above the Lee-Yang formula (for non-negative potentials, the Born series is an
alternating series with~$a\le a_0+a_1+a_2$).
Had we chosen a wave-function with~$\bgrundc
\hat{b}_{\vec{0}}^{\phantom\dag}\hat{b}_{\vec{0}}^{\phantom\dag}\bgrund=0\,$, the first three
terms in eq.~(\ref{eq:h2}) would be missing. The leading order in eq.~(\ref{eq:newbound})
would then read~${2\pi\hbar^2(a_0-a_1+a_2)}n/m$ which because of~$a_1<0$ lies above our
bound. For physically reasonable densities of up to~$128(na^3)^{1/2}/(15\sqrt{\pi})\approx
0.03$~\cite{Leggett01} and a potential such that the Born series converges fast
enough, we have~$\mbox{$(a_0+a_1+a_2-a)/a$}\ll 11.3(na^3)^{1/3}\approx 0.4$. Thus, with respect to the
exponent of the next to leading term, our bound constitutes a significant improvement
of the upper bound~(\ref{eq:bound_lieb}).

\section{Ground state energy beyond the Bogoliubov approximation} 
 We note that the Bogol\-iubov transformation exactly diagonalises the
Hamiltonian~$\hat{H}_0\simeq\hat{H}_{\rm I}+\hat{H}_{\rm II}$ (see {\it e.g.\/}
ref.~\cite{Weiss03}):
\begin{eqnarray}
 \hat{H}_0
  &=&
\frac N2\widehat{U}(\vec{0})N
+\sum_{\vec{k}}\frac{\hbar^2k^2}{2m}\akd\ak
+
\frac N2\sum_{\vec{k}\ne\vec{0}} \nonumber
\left\{\widehat{U}(\vec{k})(\akd \akmd+\ak \akm)
+2\widehat{U}(\vec{k})\akd\ak\right\}
\\
&=&Ne_0^{(0)}+\sum_{\vec{k}\ne\vec{0}}\varepsilon(\vec{k})
\hat{b}^{\dag}_{\vec{k}}\hat{b}^{\phantom\dag}_{\vec{k}}
\label{eq:hdiagonal}
\end{eqnarray}
with quasi-particle energies~$\varepsilon(\vec{k})$ given by eq.~(\ref{eq:eps}) and
 a ground state energy per particle~$e_0^{(0)}$ given by the Brueckner-Sawada-formula~(\ref{eq:e0bs}).
We now employ~$\hat{H}_0$ as the starting point for perturbation theory,
\begin{equation}
\label{eq:dec}
  \hat{H}=\hat{H}_0+\left(\hat{H}-\hat{H}_0\right)
\end{equation}
and again assume an inter-particle interaction potential such that
the Born series rapidly converges. Then already the
first order correction to the ground state energy ({\it cf.} eq.~(\ref{eq:newbound})) 
\begin{eqnarray}
e_0^{(1)}&=&\frac1N\bgrundc\hat{H}\!-\!\hat{H}_0\bgrund\nonumber
=
  \frac{2\pi \hbar^2 a_2}{m} n
+\frac{2\pi \hbar^2 a_1 }{m} \frac{32}{3\sqrt{\pi}} n\sqrt{n a_0^3}
\end{eqnarray} 
is small. 

Analogously, it can be shown that in second order perturbation theory we have
%
%
%
%
\begin{equation}
 e_0^{(2)}=\frac{2\pi \hbar^2 a_3 }{m}n +\ordn\;
\end{equation}
 as the leading order term. Thus, the terms of the Born series which are missing in
 eq.~(\ref{eq:e0bs}) are recovered, for low densities, by Rayleigh-Schr\"odinger
 perturbation theory, {\it i.e.\/}  they stem from contributions to the Hamiltonian which are
 neglected in the usual Bogoliubov approximation. Analogously, the ground state wave function can be
 modified to obtain an improved upper bound.


\section{Conclusion }
Using a quantum-mechanical variational ansatz, we have derived an upper
bound~(\ref{eq:newbound}) on the ground state energy which lies below the
bound~(\ref{eq:bound_lieb}) for physically reasonable densities if the Born series converges
fast enough. We have shown how our approach could in principle be used to further improve
this bound. However, more important than an ever improved upper bound is the fact that that there
are strong indications that a  perturbative treatment of the ground state
energy starting with the decomposition~(\ref{eq:dec}), actually 
converges towards the Lee-Yang formula~(\ref{eq:e0}) (if both the Born series
and~$a^{3/2}=a_0^{3/2}(1+a_1/a_0+a_2/a_0+\ldots)^{3/2}=a_0^{3/2}(1+\frac32a_1/a_0+\ldots)$ converge).

\acknowledgments

We would like to thank M.~Holthaus for his continuous support.


\end{document}